\documentclass[sigconf, nonacm=true, screen=true]{acmart}
\RequirePackage[utf8]{inputenx}
\input{ix-utf8enc.dfu}

\begin{document}

\author{Christopher Liscio}
\email{clliscio@uwaterloo.ca}

\author{Daniel G. Brown}
\email{browndg@uwaterloo.ca}

\affiliation{
  \institution{David R. Cheriton School of Computer Science, University of Waterloo}
  \city{Waterloo}
  \state{Ontario}
  \country{Canada}
}

\begin{abstract}
Popular musicians often learn music by ear. It is unclear what role technology plays for those with experience at this task. In search of opportunities for the development of novel human-recording interactions, we analyze 18 YouTube videos depicting real-world examples of by-ear learning, and discuss why, during this preliminary phase of research, online videos are appropriate data. From our observations we generate hypotheses that can inform future work. For example, a musician's scope of learning may influence what technological interactions would help them, they could benefit from tools that accommodate their working memory, and transcription does not appear to play a key role in ear learning. Based on these findings, we pose a number of research questions, and discuss their methodological considerations to guide future study.
\end{abstract}



\title[Watching Popular Musicians Learn by Ear]{Watching Popular Musicians Learn by Ear: A Hypothesis-Generating Study of Human-Recording Interactions in YouTube Videos}

\maketitle

\section{Introduction and Motivation}
\label{scrivauto:6}

Instrumentalists that play popular music are often faced with the challenge of learning to play melodies, harmonies, complex solos, and entire songs \textit{by ear}. These musicians use recordings in place of sheet music, and they interact with those recordings as they work towards playing the music they hear on their instruments. With the growing popularity of music streaming \cite{IFPI2022}, these interactions now frequently occur on smartphones and computers using the software supplied by the vendors of music streaming services. Aside from a near-unlimited library of music, these apps offer little beyond the set of \textit{human-recording interactions}---one's ability to influence the playback of a recording---that are provided by record, cassette tape, and compact disc (CD) players.

Purpose-built hardware and software products offer additional interactions with recordings that musicians may find helpful while learning by ear. They offer features such as placing marks at precise locations in the recording, repetitive playback between set locations (i.e. looping), and the ability to control the rate and pitch of playback independently. Despite the existence of such products, there is little research into the use of these new interactions, and how they benefit musicians as they learn to play music by ear.

To study musicians' interactions with recordings---in hopes of improving upon them, or designing novel technology that enables new ones---we must \textit{observe} musicians as they learn music by ear. Merely \textit{asking them} to describe their interactions with recordings is unlikely to yield results; musicians appear to have trouble explaining how they learn by ear when asked \cite{Bennett1980a, Green2017a}.  Unfortunately, collecting such observations would be difficult, and somewhat risky. For starters, many musicians prefer to learn by ear in private \cite{Bennett1980a, Green2017a, Groce1989}. We also wish to observe those musicians who have experience learning by ear, because they can demonstrate effective strategies. To do so would incur travel, equipment, and recruitment costs that would be prohibitive, not to mention a significant use of both the researchers' and participants' time. Also, such a study would certainly require ethics review, placing yet another barrier between the would-be researcher and this pilot study. At such an early stage, we strive for an approach that presents the least risk in these regards.

Before we commit to the expense of such a study, and to account for a dearth of foundational literature, we instead decided to conduct this preliminary, hypothesis-generating study using qualitative evaluations of YouTube videos. Using an unsophisticated querying and filtering strategy, we obtained a set of data that resembled that of a hypothetical study where a number of sufficiently-qualified musicians were recruited, and asked to film themselves learning music by ear while describing their process. The set of videos were imperfect, though one might expect such results for a similarly open-ended study. Still, we managed to uncover a number of hypotheses about the ear-learning task and how it connects to technology, which are grounded in real-world data, and we gained valuable insights that will help design future studies that could involve more conventional methods.

While the videos we analyzed all contained single musicians, some discussed how this activity prepared them for performance alongside other musicians. However, we did not observe musicians discussing how preparing for collaborative performance impacts the ear-learning process. In future experiments we will explore this further---the connection between individual ear learning sessions and ensemble performances.

Our contributions include: (1) a collection of hypotheses and research questions based on real-world observations of musicians learning by ear, and (2) a design for preliminary studies of activities that are otherwise difficult to observe, but for which video evidence may be found on YouTube or similar online video services.

We discovered some remarkable elements among the 18 videos we studied that invite further inquiry. Few videos featured musicians that learn the entirety of songs, and evidence suggests this may result from repetition in popular music: by learning to play the first verse, the musician also knows the second. Perhaps this repetition could be exploited to facilitate the process more quickly. We also see some musicians dedicating a considerable amount of time to prepare staff notation or tablature on their computers, yet none referenced it while playing what they learned. This could suggest that transcription and playing by ear are discrete activities that just happen to occur in tandem, or that transcription in some way prepares musicians to successfully learn by ear.

Quite unexpectedly, we obtained such information-rich video data using a rather unsophisticated method of video retrieval, which we will discuss in future work.

\section{Background}
\label{scrivauto:8}

\subsection{Popular Musicians and Informal Learning}
\label{scrivauto:9}

While many genres of music are popular---at some point in time, or to some group of people---here we consider musicians playing instruments in one of the myriad genres of music that (roughly) derive from the blues originating in the southern United States---country, rock, pop, jazz, hip hop, R\&B, funk, and so on. Bennett, studying local musicians as they progressed from amateurs to professionals, defined his use of \textit{rock} to be similarly broad, claiming ``a precise definition is actually unattainable.'' \cite[xvii]{Bennett1980a}

Given the challenging nature of such a definition, and ongoing evolution of music, we wish to consider a more relaxed one: popular music is that which an aspiring musician is unlikely to find among the printed materials offered by their school or private music teachers, yet is popular among their contemporaries. In rare cases where the material may appear in the curriculum, it is unlikely to take its true form, as Campbell \shortcite{Campbell1995} notes: ``Rock music that `makes it' into a school program is thus often antiseptic, a pale imitation of its true colours.'' Sufficiently motivated  musicians have little choice but to become self-sufficient when their needs are unmet at school.  We can imagine their exasperation: ``Fine! If I can't learn what I want at school or in music lessons, I'll figure it out on my own.''

And that's precisely what they do: the by-ear learning of popular music is most often a solitary activity \cite{Bennett1980a, Green2017a, Campbell1995}. Bennett recognized that a musician's initial attempts to learn from a recording happen in private, freeing them from the worry that others may deem their skills inadequate \cite{Bennett1980a}. While Bennett characterized this as an incomplete exercise---a precursor to the song getting ``worked up'' as a group---Groce \shortcite{Groce1989} and Campbell \shortcite{Campbell1995} discovered  musicians that instead learn parts in full before practice, so that everyone came prepared. Despite these differences in how much players were expected to know before turning up to their group practice sessions, these authors all make the same claim---by-ear learning happens alone, behind closed doors.

Those musicians who spoke with Groce and Bennett engaged in \textit{song getting} to perform covers with their local bands, but Bennett found this exercise also helped develop the musical vocabulary that eventually led to the development of original songs \cite{Bennett1980a}. Green's interviews with popular musicians during the 1990s built on this research by focusing more closely on the learning process itself, and discovered that---regardless of their exposure to a formal music education---popular musicians rely upon informal, by-ear methods to develop the skills for the genre they go on to (re)produce professionally \cite{Green2017a}.

\subsection{Learning By Ear, from Recordings}
\label{scrivauto:11}

Lilliestam \shortcite{Lilliestam1996} found little research on the \textit{practice} of playing by ear, partly due to the dominance of Western art music, in written form, and the associated pedagogy in the music literature. At present, little has changed in this regard---especially when considering the specific needs and practices of popular musicians.

One of Bennett's key insights is that popular musicians use \textit{recordings} as the formal notation system from which they develop a repertoire, and ultimately their own playing technique \cite{Bennett1980a, Bennett1983}. Even when popular musicians turn to a teacher, or refer to notation while they are learning, the teacher will have learned the song by ear, and the notation was transcribed directly from the recording, as Bennett notes.
\begin{quote}[{\ldots}] rock sheet music is itself derived from recordings in most cases, and although it is transcribed by experts into the conventions of traditional musical notation, the process differs little from the direct song-getting process which I have described. The generally poor repute in which rock sheet music is held among rock musicians is inherent in the limitations of the traditional notation system: Rock musicians tend to play in ways for which conventional notation does not exist.  \cite[142-143]{Bennett1980a}\end{quote}

While on-paper notation systems---on a traditional staff, or in tablature form---can serve as a memory aid or a written form of communication with others, the recording stands as the source of truth for popular musicians. The musicians interviewed by Bennett \shortcite{Bennett1980a} learned almost exclusively from recordings heard on the radio, or during private sessions spent with vinyl records or cassette tapes. Similarly, the musicians interviewed by Groce \shortcite{Groce1989} were given cassette tapes from which to learn their individual parts before band practice, and the young rock musicians interviewed by Campbell \shortcite{Campbell1995} interacted with recordings played from either cassettes or CDs.

\subsection{Purpose-Built Technology}
\label{scrivauto:13}

A musician's ability to interact with a recording is limited by the technology used to play it. At one extreme, a radio offers the least control. Musicians encounter songs by chance, and must await the next opportunity to hear it again---ideally with their instrument in tow. Once musicians obtained physical copies of a recording, record players afforded them the ability to start playback at will. This level of control is essential, because ``recorded songs are not gotten through the usual mode of audience exposure to playback events, but by the specifically defined event of copying a recording by playing along with it and using the technical ability to play parts of it over and over again.'' \cite[p. 138]{Bennett1980a}

If their turntable allowed it, the musician could also slow the playback of a 33rpm record to 16rpm with the press of a button. This interaction allowed musicians to hear quickly-played phrases at half speed. However, some musicians appeared to require more control than this, and used their ingenuity. As Jerry Garcia of the Grateful Dead recounted in an interview with Bill Barich, ``I'd picked up the five-string banjo in the Army. I listened to records, slowed them down with a finger, and learned the tunings note by note.'' \cite{Barich1993}

Fortunately, after digital audio and DSP capabilities were commonplace, such control over recordings became easier. Hardware products like the TASCAM CD-GT1, the Amazing Slow Downer\footnote{\href{https://www.ronimusic.com}{https://www.ronimusic.com}} software, and even the YouTube web player can modify the speed of playback without affecting the pitch of the recording. These features are made possible using classical approaches like waveform similarity overlap-add (WSOLA) in the time domain and the phase vocoder that operates in the frequency domain \cite{Driedger2016}.

Many more technology products and features were purpose-built to provide additional recording interactions like the above that could prove useful for musicians learning by ear. For example, automatically looping segments of audio, or placing markers to indicate the start of a verse or chorus. Such features and interactions are what we aim to study and develop further in future work.

\section{Related Work}
\label{scrivauto:15}

\subsection{Ear Learning Methods}
\label{scrivauto:16}

Previous attempts to study ear learning have largely focused on participants with little to no professional musical experience: high-school or university students enrolled in (possibly private-sector) one-on-one or ensemble music lessons, or non-musical novices. Lahav et al. recruited musically naïve college students in their study of auditory-motor interactions as the students learned melodies by ear on a piano \cite{Lahav2005}. Varvarigou \& Green categorized the ear learning styles and strategies from in-lesson observations of 75 music students, using the students' initial contacts with isolated recordings of bass melodies to characterize their learning styles, and subsequent interactions to identify strategies \cite{Varvarigou2015}. McPherson et al. conducted interviews with high school wind instrumentalists after administering ear learning proficiency tests to ask them about their approach to the task \cite{McPherson1997}. Oswald studied the methods used by high schoolers to learn melodies by ear using custom-built software that was instrumented to measure the frequency of their interactions with the music \cite{Oswald2022}. Few studies have focused exclusively on the techniques employed by experienced players. Woody \& Lehmann recruited 24 college musicians with both formal and informal experience to learn melodies by ear, and reported their strategies based on post-activity interviews with these musicians \cite{Woody2010}. Johansson instead studied the by-ear chord learning strategies by observing and interviewing musicians with far more experience---having played an average of approximately 20 years each---with even representation among six (reportedly all-male) bass, keyboard, and guitar players trained either informally or formally \cite{Johansson2004}.

Many ear learning studies have their participants learn only short melodic phrases. Lahav et al. \shortcite{Lahav2005} had participants learn the melodies from eight-bar songs using custom-designed software that synthesized MIDI notes on virtual instruments, allowing them to hear the accompaniment alongside the melody they learned to play by ear. Oswald had students also learn eight-bar melodies, though they were played from solo recordings using custom-designed audio software \cite{Oswald2022}. Varvarigou \& Green allowed students to listen to a repeating four-bar ``pop-funk style'' pattern played by a full band, but the students learned the bass melody while listening to a solo recording of it \cite{Varvarigou2015}. In contrast, Johansson \shortcite{Johansson2004} only presented participants with full band recordings while studying the chord learning strategies of experienced rock musicians, though they were asked to play along with the recording and learn as they were hearing it for the first time.

\subsection{Studying Human Behaviour on YouTube}
\label{scrivauto:18}

Our study is not unique in that we use videos posted to YouTube as a proxy for in-person observations of what people do ``in the wild'', and we are certainly not the first to do so in the HCI literature. While we have not found prior work examining online videos of musicians learning music by ear, the diverse set of human activities for which YouTube videos proved fruitful reinforces our choice. Observed activities in HCI include cooking \cite{Paay2015}, use of touchscreens \cite{Vatavu2022}, and the configuration of multi-modal game inputs \cite{Wentzel2022}, while in pediatric health care we find researchers measuring pain levels using videos of infants receiving immunizations \cite{Harrison2014}.

\subsubsection{Interactions with Technology}
\label{scrivauto:20}

YouTube provides researchers with access to a rich source of qualitative research data that might otherwise be difficult to collect. For instance, Paay et al. studied how people utilized kitchen space while cooking together, and claimed that placing a researcher and/or camera in the homes of participants would be impractical, and detrimental to such a study \cite{Paay2015}. Not only do YouTube video studies allow researchers to observe people in their homes or workplaces, but there is potential to study larger sets of data. For example, Mauriello et al. \shortcite{Mauriello2018} analyzed 1,000 YouTube videos to study how novices use thermal cameras. In other HCI studies, the use of YouTube videos gave researchers direct access to observe how people with a range of physical disabilities interacted with various technologies like touch screens and game controllers \cite{Anthony2013, Wentzel2022, Vatavu2022, Goncalves2023}. For example, Wentzel et al. \shortcite{Wentzel2022} performed a content analysis of 74 YouTube videos to identify how multi-modal inputs were configured to control PC and console games. It would be challenging to execute such a study in person, let alone find enough willing participants that fit the study's complex needs.

\subsubsection{Outside HCI and HRI}
\label{scrivauto:22}

The use of YouTube in qualitative studies extends to other fields, like health and medicine. Studies in public health have used qualitative analysis of YouTube videos to study online (mis)information about viruses \cite{Basch2015, Pathak2015, Basch2017, Basch2020}. Kong et al. \shortcite{Kong2019} studied YouTube videos depicting 25 different vape tricks to better understand how vaping is promoted to youth online. Hawkins and Filtness \shortcite{Hawkins2017a} analyzed the content of 442 videos on YouTube to study perceptions of driver sleepiness. Madathil et al. \shortcite{Madathil2015} conducted a systematic review of content and frame analysis studies to evaluate the overall quality of healthcare information available on YouTube. Harrison et al. \shortcite{Harrison2014, Harrison2018} performed qualitative content analysis of YouTube videos to review methods used to soothe infants during immunizations and blood tests.

Many of these studies, across all fields, claim either explicitly or implicitly that one of their benefits is systematicity. We disagree, and are not the first to do so. Sampson et al. \shortcite{Sampson2013} performed a systematic review of YouTube-based consumer health studies to inform the methodology they used in subsequent video reviews \cite{Harrison2014, Harrison2018}. In their review, the researchers noted that content changes daily, and the relevance algorithm on YouTube is proprietary and unstable. They propose that one can overcome this challenge by ending the screening process based on pre-defined stopping criteria; e.g., once 20 consecutive videos are found ineligible. Sampson et al. suggest that the researcher must be comfortable knowing that videos will be missed, yet they also claim---without evidence---that ``the likelihood of missing a large number is low given the relevance ranking'' \cite[p.11]{Sampson2013}. We find this notion of systematicity difficult to reconcile with our experimental discovery that YouTube search is highly unstable: we discovered significant differences between queries spaced only minutes apart. We intend to share these findings alongside the technical and theoretical challenges of running a YouTube-based video study in future work.

\section{Study Goals and Approach}
\label{scrivauto:25}

We wish to understand how current, experienced popular musicians interact with recordings as they learn music by ear and identify opportunities to design technological tools that would improve their effectiveness. To achieve this, we must first understand how musicians currently use technology, how they control recordings as they learn, and what strategies they use to replicate the notes they hear on their instrument.

Literature focused on the process by which popular musicians learn by ear is scant. Hence, we lack a theoretical frame upon which to build and test hypotheses. Thus, we chose to form hypotheses from the notable phenomena that emerged while observing musicians learning music by ear. Our study was influenced partly by that of Rueben et al. \shortcite{Rueben2021}, who also lacked a theoretical frame, and also conducted a hypothesis-generating study to understand how participants formed mental models of a robot's behaviour. Where our method differs most notably is that we use user-generated YouTube videos, of which we have no control over their content, and enjoy no guarantees that relevant examples will be found.

We are interested primarily in musicians with experience learning by ear. They already have a set of strategies that afford them the ability to expand their repertoire in this way. However, we care less about whether these musicians are \textit{professionals}. For example, popular musicians making a living from performances of original music may rarely need to learn by ear, while hobbyists could master the skill by regularly learning new songs for their own entertainment.

Participants who publish videos of themselves learning music by ear have presumably developed sufficient competency to suit our study. Thus, we automatically exclude those who do not. This self-selecting nature of our study population is akin to recruiting participants that: (1) have established by-ear learning strategies, (2) can film themselves performing the task, and (3) can demonstrate the process clearly while also explaining their actions.

\section{Methods}
\label{scrivauto:27}

\subsection{Video Collection}
\label{scrivauto:28}

Using the YouTube Data Tools\footnote{\href{https://tools.digitalmethods.net/netvizz/youtube/}{https://tools.digitalmethods.net/netvizz/youtube/}} website to perform our queries, we combined the results from 5 searches executed between May 2--5, 2023 to generate a list of 255 total videos. For each search, the query string (\texttt{"learn songs by ear"|"learn music by ear"|"learn tunes by ear''}), date range (prior to January 1, 2023), order (by relevance), and maximum number of videos (200) was held constant. YouTube's search results performed via the official YouTube API and website both failed to produce consistent results: the order of videos would change, and dozens of videos would disappear from the list---even with queries performed minutes apart. Thus, we merged multiple queries performed using the YouTube Data Tools, and selected unique video IDs to form the final set. We imported the complete list of 255 videos into Microsoft Excel so we could apply labels manually as we reviewed their content and decided whether to admit them.

Conceptually, we treat these videos as a \textit{sample} of the corpus available on YouTube, and make no attempt to suggest our approach is systematic. Further, we view this collection of videos as analogous to responses from a call for participants. Just as recruitment may yield a number of inappropriate or unqualified interviewees, videos require scrutiny before including them in the study.

\subsection{Video Selection}
\label{scrivauto:30}

We used a filtering approach inspired by that of Nielsen et al. \shortcite{Nielsen2023} to select relevant videos. Specifically, we applied high-level labels to each video after briefly reviewing their content, and retained only English-language videos depicting genuine instances of instrumentalists learning by ear. We rejected many videos in seconds: if we failed to identify an instrument while scrubbing the timeline and reviewing video thumbnails, the video was eliminated. For example, if the video contained only a talking head or graphical slideshow, but the content still seemed relevant to by-ear learning, the video would be categorized as \textit{describing-not-doing}, and thus rejected. Such efficiencies helped make this video study tractable.

When we encountered videos depicting an instrument in the hands of a musician, they got slightly more scrutiny---we sampled brief segments of those videos to assess whether the player was \textit{legitimately learning} the material in an audio recording, or merely giving a \textit{prepared lesson}.\footnote{We analyze prepared lesson videos in a study completed after the present one \cite{Liscio2024}.} For example, one excluded video contained only hypothetical examples based on nursery rhymes, and the presenters `acted out' the search for notes on their instrument. 

168 of the videos in the set were uploaded to the same channel, and largely depicted musical performances or comedic content. The musical performances were given by a solo pianist, but the comedic videos were entirely unrelated: they featured animated musical performances from popular movies and TV shows, with the original soundtrack replaced by the sound that would be heard if the animated character struck the notes they appeared to play. Fortunately, these two categories of video from this channel used a consistent title scheme that allowed us to apply labels \textit{en masse }based on the video's metadata. One video from this uploader claimed to demonstrate how they learn by ear, however it was a six hour long livestream. While sampling short intervals of this video, we found instances of the player taking requests and performing for their audience, and the video was excluded.\footnote{Later in the study, we discovered this video contained legitimate segments of learning that we missed. Had it passed the initial screening, we would have likely excluded it based on its six-hour duration.}

Three of the videos in the collection were segments of a long transcription session, and we decided to exclude the last two. This move echoes a strategy we found among the findings from Sampson et al.'s \shortcite{Sampson2013} systematic literature review of YouTube studies: omitting all but the first in a multi-part series.

This filtering process yielded a total of 18 videos for further analysis, which we labelled V1 through V18, and will make available on request alongside the identifiers of excluded videos.

\subsection{Overview of the Videos}
\label{scrivauto:32}

The 18 videos we chose to study ranged from a minute and 31 seconds in duration to over an hour and 14 minutes; the average was approximately 22 minutes, half the videos were shorter than 15 minutes, and their combined duration was six hours and 38 minutes. According to the metadata, videos were uploaded to YouTube between November 4, 2017 and November 18, 2022.

Two videos depicted saxophonists, two depicted pianists, and the rest depicted guitarists. All 18 videos in our collection depicted perceptibly male presenters.

Overall, these videos failed to garner a large audience. The most-watched video had 28,923 views, and the median view count was only 541. For perspective, one video in the original set of 255 was watched more than 8 million times. Videos with only a few dozen views carried some of the most valuable footage.

While viewing these videos, it often felt like watching responses submitted to a remote participant study. Had we requested musicians to film themselves learning a piece of music by ear while talking us through their process, we would expect to obtain a set of videos that closely resemble those we found on YouTube. However, not all the videos were as transparent and \textit{raw} as we would like.

For example, two of the videos---V6 and V8---came from the same source in \textit{livestream} format, where the guitarist interacted with a live audience via text chat. While this video appeared to be unedited, and the musician in the videos seemed adept at learning songs by ear, their behaviour in the video was clearly influenced by the virtual presence of an audience. For example, they learned songs by request, and at one point they got stuck on a chord and exclaimed, ``Gosh darn it. This is what I was worried about. Now I'm going to be stumped here.'' This may have been an expression of embarrassment and/or discomfort with struggling in front of others, and lends credence to Bennett's assessment regarding the desire for privacy during this learning activity \cite{Bennett1980a}.

We also encountered heavily edited videos, such as V4, which was only 91 seconds long, and presented disjoint segments of the ear learning process. Despite this brief presentation, the remaining footage demonstrated genuine learning: we observed the pianist making mistakes until they found the correct notes. However, most edited content was not so succinct. Rather, editing was used to intersperse recordings of the presenter's computer screen.

\subsection{Video Analysis}
\label{scrivauto:34}

The first author watched each video in its entirety while taking rough notes with the footage. As this is a preliminary study, and videos often contained stretches of irrelevant content, our notes varied in their level of detail. Our goal was not to transcribe the videos, but rather to collect a mixture of high-level summaries, timestamped quotations, and brief descriptions of notable events from each video.

The authors met regularly to discuss remarkable findings that emerged from the videos, and those deemed worthy triggered further study. Videos were reviewed over the time ranges relevant to the phenomena, paying close attention to different details each time. For example, once we deemed it significant that musicians often sang melodies, we re-watched those videos, using our notes to direct us to relevant segments. As we watched, we looked more closely to note whether they sung alongside the recording, after stopping playback, or while identifying notes on their instrument.

Late in the study, these reviews became more frequent as we continued to refine our findings. It grew cumbersome to move between related segments across many videos using YouTube's web player. Limitations in network speed, YouTube's playback controls, and awkward navigation between browser tabs made progress difficult. To improve our efficiency, we downloaded the 18 videos to local storage, and loaded them into Final Cut Pro X. There, we assigned keywords to video segments, and could watch related segments from different videos in rapid succession.

\subsection{Summary}
\label{scrivauto:36}

The above method of querying, filtering, and analyzing content of videos is suitable only for a preliminary, hypothesis-generating video study like ours. For instance, our query strategy could have mirrored HCI studies that maximize the number of relevant videos using a combination of keywords to form a larger set of search terms (e.g. \citealp{Anthony2013, Wentzel2022, Vatavu2022}). However, breadth was less of a concern for us at this stage, because we could not yet handle a large quantity of video material. Instead, we will reserve more sophisticated query strategies for future work that tests our hypotheses, provided online video content is the appropriate medium to do so.

\section{Results}
\label{scrivauto:38}

Here, we present common patterns that were identified during our analysis of the video footage.

\subsection{Scope of Learning}
\label{scrivauto:40}

In only three of the videos did we observe musicians learning to play songs in their entirety. That is, these musicians worked towards playing their instrumental parts for the full duration of the recording. Among them, only the guitarist in V2 provides us with evidence of his success---performing the song he learned at the end of the video.

In the remainder of the videos, we find musicians learning only \textit{portions} of songs---solos, riffs, or a subset of the chords. For example, the guitarist in V11 learns the riff at the very start of a recording, in V16 only the solo is learned, and the guitarist appearing in both V6 and V8 learns chords within excerpts of songs that were requested by his audience.

\subsection{Transcription and the Role of Notation}
\label{scrivauto:43}

In three of the videos (V2, V17, and V18), we observed musicians transcribing notes from the recording to produce sheet music for the songs as they learned them. Unlike the process of transcribing speech, wherein words are written down as they are recognized, the musicians in these videos did not record the notes they heard until they were located on their own instrument.

While the guitarist in V17 and the pianist in V18 stated explicitly that their video's goal was to demonstrate the transcription process, only the guitarist in V2 explained why he generates tablature: the notation helps him reason about rhythm patterns, and serves as a memory aid. However, he contradicts this statement by learning the song's (quite rhythmically complex) solo later in the video without appearing to create or use any tablature.

Musicians recorded notation onto a staff (V18), or as guitar tablature (V2 and V17) using software designed for those purposes. To verify the correctness of their sheet music, the presenters in these videos did not appear to read what they entered. Instead, they used the software's built-in synthesizer to perform their sheet music virtually so they could assess whether the notes were representative of the original recording.

\subsection{Use of Technology}
\label{scrivauto:45}

It was often unclear what technology the musicians used to play and interact with recordings, but in eight of the videos we could identify it visually. Musicians used YouTube for music playback in three of the videos, three used non-specialized players (the Music app on iPhone, Spotify, and iTunes), one used Digital Audio Workstation (DAW) software, and only a single video featured purpose-built Mac software called Transcribe!\footnote{\href{https://seventhstring.com}{\href{https://seventhstring.com}{https://seventhstring.com}}}. Interestingly, no musicians used dedicated hardware devices---they used software running on a smartphone or computer.

Because many videos did not show technology prominently in the frame, we tried to infer the use of purpose-built \textit{features} using a combination of body language, dialogue, and apparent changes in audio playback. For example, in V12 the saxophonist's shoulder raises slightly before stating YouTube was used to slow playback, we hear the music start, and his shoulder lowers again. We judge this to indicate that he started playback of the recording at a slower rate than the original. Despite our increased sensitivity to detect off-camera use of purpose-built technology features, we found that only four of the 18 videos contained evidence of their use. In three of these, musicians \textit{slowed audio playback} with no change in pitch. One of them---the guitarist in V17---also used Transcribe!'s \textit{looping playback} to help identify a finger-picking pattern. The guitarist in V1 did not slow playback, but appears to have used \textit{marked positions} to start playback repeatedly from the same spot in the recording.

\subsection{Temporary Note Retention}
\label{scrivauto:47}

Eight of the videos contained examples of musicians singing (or humming) the notes they heard in the recording; most did so while the recording was stopped. Those who continued to sing notes as they sought matching pitches on their instrument produced audible \textit{goal tones,} and worked to minimize the error between their instrument and their voice. Others seemed to repeat notes with their voice as if to hold them in memory; just as one might recite a phone number before writing it down.

In eleven of the videos, musicians played notes on their instrument while the recording continued to play. It sometimes \textit{appeared} that the musician was looking for notes while playback continued; not relying on their memory. However, on closer inspection we noticed this was not always the case. Some musicians played phrases heard moments earlier; without pausing the recording, they could still recall those notes. In those instances where the musician did not appear to remember what was just heard, they were looking for \textit{anchor notes}: to help identify chords by bass note, or find the key of the song. For example, the saxophonist in V12 played scales over the recording to identify which was used by the soloist.

There were ten videos in which musicians demonstrated they could remember notes shortly after hearing them. Sometimes they repeated the notes vocally, but we could only hear notes emitted from the instrument while the musician looked for the ones they had in mind. We see an example of this in V12: the saxophonist sings the phrase alongside the recording, repeats it after the recording is stopped, then attempts to replicate the phrase on his saxophone.

\subsection{Familiarity with the Music}
\label{scrivauto:49}

In eight of the videos we analyzed, musicians claimed explicitly that their on-camera attempt to learn the recording was their first. Sometimes, it was their first time hearing the song, as promised by the guitarist in V11. However, in V14 the guitarist was learning a song that was released on the day of filming, and he claimed that before the video was recorded he listened repeatedly to the song while visualizing himself playing notes on the fretboard; little time elapsed before he replicated the introductory riff in the video.

While musicians may have heard recordings for the first time in their videos, they seemed to be familiar with similar music. For example, the guitarist in V10 claimed, ``I don't even know this one'' while listening to one of the songs learned in the video, but claimed he knew other songs by the same artist; he could play each of the country songs in the video shortly after hearing them. Similarly, we see a rock guitarist in V14 learning a rock song, and a metal guitarist in V2 learning metal. While the guitarist in V7 appeared to be familiar with jazz music, he appeared to struggle more than others to identify jazz chords in the recording, though they were originally played on a piano.

\subsection{Application of Music Theory}
\label{scrivauto:51}

Among the set of videos, eight guitarists and one pianist worked to identify chords by name. Once found, they named a root note, whether the chord was major or minor, and any extensions or inversions where applicable. The methods used to identify chords differed between musicians. For example, in V13, the bass notes are identified first, and the guitarist auditions major and minor variations to identify which sounds correct---a trial and error approach. By contrast, the guitarist in V6 uses music theory: he first identifies the song's key as F major, then uses his knowledge of (diatonic) chords to readily name chords numerically (e.g., ``the four chord'' (IV), ``the two chord'' (ii)). Drawing from a vocabulary of common chord progressions (e.g., vi-ii-V-I, I-IV-V), this guitarist could also identify \textit{sequences} of chords.

We find a similar dichotomy in the methods used to learn individual notes: some musicians did not seek notes until the song's key was identified, whereas others \textit{hunted} for their location on the instrument. For example, the guitarist in V11 chooses an arbitrary note, then moves his fingers to adjacent frets in the direction that corresponds to his perceived difference: moving in the direction of lower pitches when the note is too high, and vice versa. By contrast, once the saxophonist identified the correct pentatonic scale to play in V12, he only produced notes within the scale and appeared to mimic the phrase rather quickly.

For those that sought the song's key, there was also a mixture of techniques that we observed. In one, the musician looks for \textit{home base}: a chord or note that stands out, or ``feels like home'' according to the guitarist in V15. That is, they look for a single chord or note that sounds like it could serve as the root. This method is sometimes used in concert with trial and error, wherein the musician auditions candidate \textit{scales} (e.g. pentatonic, diatonic, or a relative minor) to identify the key.

\section{Discussion}
\label{scrivauto:55}

In \S\,\ref{scrivauto:40} we observed few musicians learning to play songs in whole; the majority were focused on solos, riffs, or a portion of the song's chords. This could be a consequence of the repetition found in pop music \cite{deClercq2011}: a musician that learns one instance of a chord progression, verse, or chorus, can play the others in the song. However, experienced musicians might only learn by ear when they encounter novel segments; to mimic a certain technique, or copy a challenging passage. Of course, it is plausible that these videos are kept short to suit the medium: YouTube limits uploads to 15 minutes in duration for unverified accounts\footnote{\href{https://support.google.com/youtube/answer/71673}{https://support.google.com/youtube/answer/71673}}, and succinct videos may attract more viewers.

Our observations in \S\,\ref{scrivauto:43} suggest that while by-ear learning is necessary to produce notation, the converse is not true. Specifically, we saw musicians learning music by ear without writing anything down or producing notation. Our findings further the claim that recording notes on a staff, as tablature, digitally, or on paper, serves only as a memory aid or transmission mechanism for the material that was originally learned by ear \cite{Bennett1983}.

We saw in \S\,\ref{scrivauto:45} that most of the musicians did not employ purpose-built technology features. As a result, it appears they are no better off than their counterparts were 40 years prior. Thirteen musicians appearing in 14 of the videos\footnote{The same musician appears in V6 and V8.} only listened to recordings at full speed and were content to cue playback of passages with little precision. These musicians could be handed a record, cassette, or CD player loaded with the same music they learned in their videos, and---provided they could operate such archaic equipment---the learning activity would be largely indistinguishable.

The musicians we observed in \S\,\ref{scrivauto:47} used different strategies to transport the notes heard in the recording to their instrument. They often used their voice: an instrument on which they can reproduce the notes they hear more readily. However, if they stopped singing the notes, and the recording was left paused, these musicians relied upon their\textit{ mind's ear} \cite{Covington2005}: they could mentally retain notes of interest, and locate them on the instrument without the need for supplementary audible feedback. This skill seems necessary for instrumentalists---e.g., trumpeters, saxophonists---whose breath is required to produce sounds on the instrument, and vocalizing notes while also playing them may be impossible.

When musicians claimed they were learning (or hearing) songs for the first time (\S\,\ref{scrivauto:49}), their statements did not appear to characterize typical learning sessions. Rather, we interpreted these as signals of the musician's intent to portray a \textit{genuine attempt} on camera, and establish credibility with their audience. While musicians might learn new material on short notice---to substitute for another instrumentalist (V1), or learn songs on the spot for their students (V15)---it seems beneficial to develop familiarity beforehand to make learning go more smoothly (V14). When that is not possible, musicians can draw upon their familiarity with the genre, or follow common practices on their instruments to make learning more efficient. For example, the musicians in V2, V10, and V14 could readily draw upon guitar-oriented idioms, but the guitarist in V7 had to discover the \textit{voicing} on the guitar that matched the piano in the recording; he worked more slowly than the others. This guitarist did not demonstrate the use of established \textit{finger routes}: the shapes and scales that are easily recalled, and set the frame of what the instrumentalist can play \cite{Lilliestam1996}.

We reported the use of diverse methods for identifying notes and chords in \S\,\ref{scrivauto:51}, which call upon varying degrees of music theory, and include some of the chord finding strategies identified by Johansson \shortcite{Johansson2004}. None of the methods we observed were \textit{theory-free}---merely \textit{naming} chords and notes forms a link between what is heard and the formal language of music---but those musicians who applied theory to their process could more readily identify notes and chords than the others who employed trial-and-error methods. For example, once the key is identified, the musician has only 7 of 12 notes to consider, knowing the key's diatonic chords indicates which are major or minor, and familiarity with pop music harmony shrinks this list further \cite{deClercq2011}. However, we suspect that one's perceived knowledge of music theory might be a proxy for their instrumental proficiency. For example, a saxophone player who practiced a scale hundreds of times has ostensibly learned its name, can play it comfortably, and may recognize the sound of its intervals. Similarly, a guitarist with a sizeable repertoire has played many chord progressions, and may anticipate---or hear---the most common sequences of chords. Thus, it seems unlikely that an understanding of music theory in the absence of instrumental skills would facilitate the rapid playing songs by ear.

\subsection{Hypotheses and Future Work}
\label{scrivauto:57}

Our findings raised a number of questions that, when answered, will provide us with insights to guide the design of purpose-built technology and identify novel human-recording interactions. For example, knowing that only a subset of interactions are required for a given scope of learning---an entire song, or only its solo---designers could ensure that those interactions are easily accessed in those scenarios they are needed most. Additionally, understanding the impact of a musician's working memory on their ability to copy notes from a recording could inspire the design of novel interactions that help those who struggle to do so. Here we discuss our hypotheses in further detail, and provide \autoref{tab:hypotheses} for an overview.

\begin{table*}[t]
\begin{tabular}{l c c}
\toprule
Hypothesis & Results & Discussion \\
\midrule
Desiderata recording interactions change depending on learning scope & \S\,\ref{scrivauto:40} & \S\,\ref{scrivauto:60} \\
Transcription serves no role in the by-ear learning task & \S\,\ref{scrivauto:43} & \S\,\ref{scrivauto:64} \\
Experienced musicians rarely engage with purpose-built tools & \S\,\ref{scrivauto:45} & \S\,\ref{scrivauto:66} \\
A musician's working memory will dictate their ability to copy notes & \S\,\ref{scrivauto:47} & \S\,\ref{scrivauto:69} \\
Intentional recording familiarization improves by-ear learning & \S\,\ref{scrivauto:49} & \S\,\ref{scrivauto:73} \\
Knowledge of theory helps make note and chord identification more efficient & \S\,\ref{scrivauto:51} & \S\,\ref{scrivauto:75} \\
\bottomrule
\end{tabular}
    \caption{Our hypotheses, their basis in our results, and the sections in which they are discussed.}
	\label{tab:hypotheses}
\end{table*}

\subsubsection{Scope-Oriented Interactions}
\label{scrivauto:60}

Musicians learn songs in part or in whole. We would like to further understand this dichotomy, and think the musician's scope of learning may dictate which human-recording interactions are helpful. On the surface, this seems obvious: surely a musician learning solos needs to slow playback, but a musician seeking chords or learning the song's structure does not. However, our videos contained evidence to the contrary. The guitarist in V16 learned a solo from a pop song, and in V17 the guitarist was learning \textit{fingerstyle }chords; but it was the guitarist in V17 that chose to slow and loop the playback, while the guitar solo was learned at full speed.

A future study can test this hypothesis by recruiting musicians that have experience learning entire songs, melodies, and instrumental solos by ear. Participants could be presented with suitable (to their instrument, genre) recorded material to learn from. Ideally, this study would occur over two phases: one where participants learn using their preferred tools, and one where we supply a tool with a wider range of human-recording interactions. In the latter configuration, the participants would be provided with sufficient training to access the entirety of these interactions; the former is necessary to control for existing habits that drive participants towards oft-used features.

A different study can test novel human-recording interactions for those musicians who learn entire songs; for example, structure-oriented graphical representations that exploit the repetitive nature of popular music. Recordings could be presented for participants using an instrumented application that presents one of three modalities: a timeline, a waveform representation, and a structural rendering of measures and sections. Each would provide navigation controls appropriate to the representation---e.g., tapping on numbered measures in a structural interface, or scrolling between locations on a timeline---and interaction data could be analyzed using appropriate metrics to report the user's efficiency.

\subsubsection{Limited Value of Notation}
\label{scrivauto:64}

Our findings suggest that notation provides little benefit even to \textit{literate} musicians (that can read music) while they learn by ear. This hypothesis could be tested by an in-person or remote study that separates participants into two groups: those free to transcribe as they learn a piece by ear, and those asked not to record anything onto paper. Because transcriptions vary based on one's chosen instrument, and the detail of the material they are learning, these results could shed light on the unique needs of certain instrumentalists. For example, sight-reading pianists might find immediate utility in their transcriptions, whereas guitarists might instead use tablature to refresh their memory during the next learning session. A similar study could replicate this same test, except they can provide purpose-built technology for by-ear learning that allows the creation and display of notation.

\subsubsection{Experience and Purpose-Built Technology}
\label{scrivauto:66}

Having observed such infrequent use of purpose-built technology features, we wonder if musicians with experience learning by ear no longer need them, or indeed if no one does. A study consisting of a survey and interviews would help researchers understand the proliferation of purpose-built tools, and their suitability for this population. Researchers could also determine if purpose-built tools are used less frequently as a musician's learning skills improve; a longitudinal study could follow early-intermediate instrumentalists to see how their preferences shift with experience.

However, we believe that there exists a set of interactions---existing, as well as novel ones---that benefit this population. Researchers can conduct a diary-like study wherein experienced musicians are provided with a full suite of purpose-built features, and training to ensure they are proficient with them. During the study period, musicians would record instances of by-ear learning with the provided technology: capturing their preparations for gigs, or session work. In addition to the footage of their interactions with the technology, the tool can be instrumented for further data capture, and the musicians would be encouraged to share their thoughts about their time spent with the tool on video, or in diary form.

\subsubsection{Memory's Impact on Copying Notes}
\label{scrivauto:69}

People with working memory limitations may need to work differently while learning by ear, and would benefit from technological supports. For example, a wind or brass player with an under-developed mind's ear cannot continuously sing notes while seeking them on the instrument; technology that can repeat the playback of phrases, or sounds notes continuously would assist these musicians.

To measure the effectiveness of such technological supports, researchers would require: (1) a mixture of participants with varying levels of by-ear learning experience, and (2) a method to gauge and compare their working memory abilities. Such a study should account for neurodivergent participants, such as those with ADHD who may have limited working memory \cite{Vassileva2001}. For each of the proposed tools, participants could be evaluated to determine whether the tools improve their learning outcomes.

For those musicians with a well-developed mind's ear, there are some research opportunities to create tools that leverage their abilities. For example, these musicians may be helped by tools that allow the navigation of recordings in musically relevant, bite-sized chunks that match their working memory capacity. Not only can researchers test the effectiveness of such tools, but it would also be valuable to develop measurements of a musician's capacity for musical information.

\subsubsection{Familiarity Improves the Learning Experience}
\label{scrivauto:73}

A musician that spends time familiarizing themselves with songs before learning them by ear appears to gain an advantage over those musicians who fail to do so. This seems obvious when the recording is not already known to the musician. However, an intentional listening practice that precedes the by-ear learning session may improve the learning outcome by cueing associations to existing memories of the song \cite{Snyder2014}. A musician's existing familiarity with a recording could be exploited in purpose-built tools; for example, recommending songs from the musician's music library with high play counts.

To explore the impact of familiarity, researchers can test participants by supplying a previously-unheard recording of music, and let them hear it a number of times before they attempt to learn it by ear. Across the study, participants could be grouped by the number of times the song is heard before their attempt. A similar experiment---ideally with the same cohort---can test additional exposure to the music by using songs that participants are well-acquainted with, but have never before learned to play (by ear, or otherwise). Researchers could not only monitor the duration of by-ear learning sessions, but also analyze recording interactions.

\subsubsection{Knowledge of Theory Improves Identification}
\label{scrivauto:75}

It appears that musicians can identify notes and chords more easily if they can apply some level of music theory knowledge. However, we are unsure exactly what the musician needs to know, or if that knowledge transcends traditional music theory. It may be less important that a musician can readily name the notes belonging to chords, and more important that they can identify common elements---chord progressions, scale patterns, etc.---that are used in the genres of pop music that they play. Woody \& Lehmann \shortcite{Woody2010} found that among college music majors, those with vernacular experience---performing in church, jazz, popular, and folk music bands---required fewer trials to learn melodies by ear, and could recognize patterns in the melodies that they described as typical or predictable.

This is worthy of further exploration; researchers should study how musicians apply their knowledge of theory during by-ear learning sessions. Once the necessary elements of theory are identified, technology can be developed to exploit and surface the relevant information that aids a musician's progress. Researchers could then test the effectiveness of their interventions, focusing not only on those musicians who have experience learning by ear and already possess theoretical knowledge, but also those who regularly learn by ear but struggle to understand music theory. Researchers might find that the latter group is impeded, and not helped by such tools.

\subsection{Limitations}
\label{scrivauto:77}

As stated earlier, the majority of musicians only shared footage of themselves learning short portions of songs. This tendency to provide a piecemeal presentation of the process makes it difficult for us to observe the strategies musicians use to work out entire pieces of music. Moreover, only one video contained a performance of the song learned by ear. Hence, we could not gauge whether the strategies we observed in the videos were actually effective.

While the self-selecting nature of musicians posting to YouTube provided videos that demonstrate competency, we risk collecting examples that are performative in nature. For example, we cannot verify that a musician has not already learned the song before filming, or that an on-camera struggle to locate notes on their instrument is authentic. Additionally, we recognize that one's behaviour is likely to change when they know they are being recorded; this is obvious in our study, because musicians would not explain their actions while learning a song alone.

The videos we studied all contained perceptibly male musicians, which is especially unfortunate. While we are aware of videos on YouTube that feature perceptibly female musicians learning by ear, our queries and filtering strategy failed to capture them in this study. In our data set, one female guitarist was excluded because she did not learn from a recording; instead, she explained how nursery rhymes can be played from memory. In another video, discovered long after our analysis, a female guitarist learns to play a solo from a recording, but the video has a title that failed to match our queries; its metadata refers to learning a specific song title by ear. We hope to rectify this imbalance by taking concrete steps in our future work to increase representation across a more diverse set of gender identities.

We intentionally used a simple query methodology in this study; it produced a small data set, and it would take little effort to expand it. However, given the nature of our study---generating hypotheses, and not presenting results based on our data---we feel this oversight is forgivable. This was a preliminary study, yet it still required considerable effort to categorize the videos, and generate data from our observations. Given the wide range of methods used to gather YouTube videos in other studies, and major differences found in repeated YouTube queries, we felt it was necessary to reserve a more comprehensive strategy for future work.

\section{Conclusion}
\label{scrivauto:84}

We set out to clarify our understanding about how popular musicians learn by ear, and observed 18 in-the-wild examples of them doing so to form hypotheses that lead us towards future studies. Using YouTube as a source of data proved fruitful for this endeavour, and allowed us to set a course for future study with minimal financial investment, and no need for ethics approval. While the review of videos was time-consuming, we made the problem tractable through the use of efficient strategies for selecting relevant videos.

Our results were obtained by comparing behaviours that we observed across the 18 videos, and led us to form six hypotheses. First, musicians learning entire songs may require a different set of human-recording interactions compared to those learning only segments of songs. Second, the act of transcribing notation does not facilitate by-ear learning. Third, purpose-built technology is not widely used by musicians who have experience learning by ear. Fourth, a musician's working memory facilities influence their ability to copy notes from recordings. Fifth, the by-ear learning experience goes more smoothly when a musician familiarizes themselves with a recording. Finally, sixth, a musician's ability to apply their knowledge of music theory helps them identify notes and chords more efficiently.

Armed with these insights, we plan to execute future studies that look more closely at this task. In future work, we hope to develop a deeper understanding of the interaction patterns employed by experienced musicians that learn by ear, and build further insight into their needs and intentions. With that, we can more confidently proceed towards the ultimate goal of expanding and improving the musician's toolbox for learning music by ear.

\bibliographystyle{ACM-Reference-Format}
\bibliography{bibliography}

\end{document}